\documentclass[superscriptaddress,nofootinbib,preprint]{revtex4}
\usepackage{graphicx}
\usepackage{amssymb,amsmath,amsfonts}
\usepackage[hypertex]{hyperref}

\newcommand{\lsim}{\lesssim}

\def\lozenge{\boxit{\hbox to 1.5pt{\vrule height 1pt width 0pt \hfill}}}
\def\ie{{\it i.e.}}
\def\eg{{\it e.g.}}
\def\etc{{\it etc}}

\def\mpl{\ifmmode M_{pl}\else $M_{pl}$\fi}
\def\mpl{\ifmmode \overline M_{Pl}\else $\bar M_{Pl}$\fi}
\def\to{\rightarrow}
\def\lsim{\mathrel{\mathpalette\atversim<}}

\def\grtsim{\,\,\rlap{\raise 3pt\hbox{$>$}}{\lower 3pt\hbox{$\sim$}}\,\,}
\def\lsim{\,\,\rlap{\raise 3pt\hbox{$<$}}{\lower 3pt\hbox{$\sim$}}\,\,}

\begin{document}

\pagestyle{plain}

\hfill$\vcenter{
\hbox{\bf MADPH-05-1432}
\hbox{\bf SLAC-PUB-11357}}
$

\vskip 1cm
\title{Off-the-Wall Higgs in the Universal Randall-Sundrum Model}

\author{Hooman Davoudiasl}
\email{hooman@physics.wisc.edu}
\affiliation{Department of Physics, University of Wisconsin,
Madison, WI 53706}
\author{Ben Lillie}
\email{lillieb@slac.stanford.edu}
\author{Thomas G. Rizzo}
\email{rizzo@slac.stanford.edu}
\affiliation{Stanford Linear Accelerator Center, 2575 Sand Hill Rd.,
Menlo Park, CA  94025}

\begin{abstract}

We outline a consistent Randall-Sundrum (RS) framework
in which a fundamental 5-dimensional Higgs doublet induces
electroweak symmetry breaking (EWSB).  In this framework of a warped
Universal Extra Dimension, the
lightest Kaluza-Klein (KK) mode of the bulk Higgs is tachyonic
leading to a vacuum expectation value (vev) at the TeV scale.
The consistency of this picture imposes a set of constraints on
the parameters in the Higgs sector.  A novel feature of our
scenario is the emergence of an adjustable bulk profile for the Higgs vev.
We also find a tower of non-tachyonic Higgs KK modes at the weak scale.
We consider an interesting implementation of this ``Off-the-Wall Higgs''
mechanism where the 5-dimensional curvature-scalar coupling
alone generates the tachyonic mode responsible for EWSB.
In this case, additional relations among the parameters of the Higgs and
gravitational sectors are established.  We discuss the experimental
signatures of the bulk Higgs in general, and those of the ``Gravity-Induced''
EWSB in particular.

\end{abstract}
\maketitle


\section{Introduction}

The Higgs mechanism in the Standard Model (SM) provides a
simple and economical explanation for electroweak
symmetry breaking (EWSB).  In the SM, it is assumed that the Higgs potential
contains a tachyonic negative mass-squared term that causes the
Higgs to develop a vacuum expectation value (vev), resulting in EWSB.
However, the source of this tachyonic mass,
which must be of order $100$~GeV, is not explained in the SM.

Another mystery regarding the Higgs sector, pointing to physics beyond
the SM, is the size of the typical Higgs mass compared to the cutoff scale;
this is the usual hierarchy problem.  For example, if we take the 4-dimensional
(4-d) reduced Planck mass $\mpl \sim 2 \times 10^{18}$~GeV to
be the cutoff scale, we would naively expect quantum corrections
to raise the Higgs mass to be near $\mpl$, which is $O(10^{16})$
too large.

An interesting explanation of the hierarchy is provided by
the Randall-Sundrum (RS) model \cite{Randall:1999ee}.
In this model, a curved 5-d (bulk) spacetime called $AdS_5$ is bounded by 4-d
Minkowski boundaries, corresponding to the geometry
of the observed universe.  The curvature of the 5-d
spacetime induces a sliding scale along the warped extra dimension,
geometrically generating the weak scale from a large Planckian scale, at
one of the 4-d boundaries.  This boundary is referred to as the TeV-brane,
hereafter.  The geometrical red-shifting in the RS model is exponential
and hence capable of explaining large hierarchies, using parameters
that are typical in the model.  In the original RS model, the
only 5-d field is the graviton{\cite{Randall:1999ee,Davoudiasl:1999jd}}.
Numerous works have since extended
the RS setup to include bulk fermions and gauge fields{\cite{Davoudiasl:1999tf}}.
However, even though the SM can be promoted to a 5-d theory, the
fundamental Higgs field
responsible for EWSB has been typically kept on the TeV-brane in the RS model.
The reason for this has been to avoid
problems associated with extreme fine-tuning{\cite{Chang:1999nh}} in generating
TeV-scale gauge boson masses which the RS model is constructed
to resolve, and conflict with known experimental data{\cite{unknown:2004qh}}
including the SM relationship $M_W=M_Z \cos \theta$.
Thus, the Higgs is treated as a 4-d field in the RS geometry. (There have also
been attempts to build RS models without
fundamental Higgs bosons{\cite{Csaki:2003dt}}
where the role of the Higgs doublet as a source of Goldstone bosons
is played by the fifth components of the gauge fields themselves.)

In this work, we study the requirements for successful EWSB, using
a {\it fundamental} 5-d Higgs doublet in the RS background.  We
show that by appropriate choices of the Higgs sector parameters in
the bulk and on the branes, one can generate a single tachyonic
Kaluza-Klein (KK) mode of the Higgs field in the low energy 4-d
theory.  This tachyonic mode is identified as the SM Higgs field.
Given a suitable quartic bulk term for the Higgs, the tachyonic
mode will lead to the usual 4-d Higgs mechanism and endow the
electroweak gauge bosons with mass.  The quartic terms will reside
on the 4-d boundaries, since they are higher dimension
operators in 5-d and are expected to be small in an
effective theory description.  Note that these terms are not
necessary for the gauge invariance of the 5-d Higgs theory and
thus we set them to zero at tree level in the bulk.
A {\it novel} feature of this
mechanism is that the Higgs vev now has a profile that extends
into the bulk and is no longer a constant.  This provides for new
model-building possibilities that we will briefly discuss in our
presentation.  A typical signature of our scenario is the
emergence of a tower of Higgs KK modes whose detection at future
colliders we will consider in this work. Since all of the SM
fields are now in the bulk, this scenario is an example of a {\it
warped} Universal Extra Dimension{\cite {Appelquist:2000nn}}.

The ``Off-the-Wall Higgs'' mechanism outlined above, like its 4-d SM counterpart,
does not explain the origin of the mass parameters of the Higgs potential.  It
would be interesting if the RS geometry itself could provide the necessary mass
scales of the 5-d Higgs sector, leading to novel connections between the
gravity and the Higgs sector parameters.  We show that a modified version of
the RS gravity sector
does indeed provide such a mass scale that can result in a successful
realization of the Off-the-Wall Higgs mechanism.
The central observation is that the most general bulk
action in the RS model should include a term $\xi R \Phi^\dagger \Phi$,
coupling the Ricci curvature scalar and the
Higgs{\cite{Giudice:2000av,Farakos:2005hz}}. Because there is a
constant curvature in $AdS_5$ as well as $\delta$-function terms on both
branes, this coupling acts as a mass term for
the Higgs and, with the appropriate choice of parameters, can lead to
EWSB. This picture provides a link between the
seemingly unrelated Higgs and gravitational parameters and eliminates
the need for {\it ad hoc} masses in the Higgs potential.

To realize this idea, we consider the most general 5-d action for
a scalar coupled to gravity, consistent with gauge symmetry and general
coordinate invariance. In addition to the
standard kinetic and coupling terms, this action then includes the
curvature-scalar coupling, along with a string motivated{\cite {Zwiebach:1985uq}}
higher curvature Gauss-Bonnet
term (GBT){\cite{Kim:2000pz}} and brane localized kinetic terms
(BKT's){\cite{Davoudiasl:2002ua}}
for the graviton{\cite{Davoudiasl:2003zt}}, as well as the aforementioned
boundary BKT's for the Higgs quartic coupling. In
principle one could directly include {\it ad hoc} tachyonic mass
terms for the Higgs. However, we find that, under the
assumption that these terms are small, there are phenomenologically
acceptable regions of parameter space where EWSB is solely
driven by the gravity sector. It is
interesting to note that in the favored region
typical values of $\xi$ are close to the conformal value $\xi=-3/16$.

This connection between
gravity and the Higgs sectors leads to experimentally observable
signals that could point to this picture as the
correct mechanism of EWSB.  For example, it is well-known that
curvature-Higgs coupling on the TeV-brane leads to radion-Higgs
mixing{\cite{Giudice:2000av}}.
The same effect exists with our bulk Higgs.  We discuss how measuring
the Higgs-radion mixing  in conjunction with other collider
measurements in the Higgs and gravity sectors can test
the gravity-induced EWSB scenario and establish its parameters.

In the next section, we will describe the ingredients for achieving
consistent bulk Higgs mediated EWSB in the RS geometry while section III
discusses the details of the generation of masses for the SM electroweak gauge
bosons.  Section IV focuses
on the gravity-induced realization of EWSB with a bulk Higgs and the
new constraints on the Higgs and gravity parameters
that need to be satisfied in a successful scenario.
Section V is devoted
to a discussion of experimental tests and the novel phenomenological
aspects of our scenario, such as the KK Higgs physics at colliders.  Here,
we also outline the measurements that will result in the establishment of
gravity-induced EWSB in the RS model. In section VI, we present our
conclusions.

\section{Bulk Higgs EWSB in RS: the General Framework}

In this section we will describe the overall framework for our model and the
general mechanism of bulk Higgs EWSB in the RS scenario. In particular we will
demonstrate how the SM gauge boson masses are generated and the appearance of
the Higgs KK spectrum. We perform our analysis using the standard
RS geometry{\cite{Randall:1999ee}}: two branes are located at the
fixed points of an $S^1/Z_2$ orbifold; between the branes, which are separated
by a distance $\pi r_c$, is a slice of $AdS_5$ and the metric is given by
\begin{equation}
ds^2=e^{-2\sigma} \eta_{\mu\nu}dx^\mu dx^\nu-dy^2\,.
\end{equation}
Here, $\sigma=k|y|$
with $k$ being the so-called curvature parameter whose value is of order the
5-d Planck/fundamental scale, $M$, and $y$ is the co-ordinate
of the fifth dimension. In order to be as straightforward
as possible and to present the essential features of the model
we will delay any discussion of the localization of the SM fermions until a later
section and for now only assume that the SM gauge and Higgs fields are in the
bulk. With this caveat our action is given by
\begin{equation}
S=S_{Higgs}+S_{gauge}\,,
\end{equation}
where
\begin{eqnarray}
\nonumber
S_{Higgs}&=&\int d^5x ~\sqrt {-g}\left[(D^A\Phi)^\dagger (D_A \Phi)\right]
-\frac{1}{k}\int d^5x ~\sqrt {-g}\\
&\times&\left\{m^2 k+[\mu_H^2 + \frac{\lambda_H}{k} \Phi^\dagger \Phi]
\,\delta(y-\pi r_c)-[\mu_P^2 - \frac{\lambda_P}{k} \Phi^\dagger \Phi]\,\delta(y)
\right\}\Phi^\dagger \Phi\,,
\end{eqnarray}
and $S_{gauge}$ is the usual action for the 5-d SM gauge fields.
Note that in addition to the bulk mass $m^2$ for the Higgs field $\Phi$ we have
allowed for mass terms on both the TeV and Planck branes: $\mu_{H,P}^2$.
The bulk quartic term for the Higgs is a higher dimension operator, thus presumably
suppressed, and is not demanded by gauge invariance in our
5-d effective theory.  We thus set it
to zero in the bulk.  However, this term is marginal on the 4-d
boundary theory and thus present with coefficients $\lambda_{H, P}$ on the
TeV and Planck branes, respectively.  Since we will be
interested in first generating
a tachyonic Higgs mass and then shifting the field as in the SM, the role of the
quartic terms is just to stabilize the potential allowing for a positive mass-squared
physical 4-d Higgs.

In order to generate EWSB, the Higgs action above must lead to (at least one) vev
in 4-d. In the absence of brane terms one may try to shift the field $\Phi$ and
then perform a KK decomposition as is usually done. This approach will not work
here for several reasons. First, we note that the absence of
a 5-d potential which could lead to a shift of the Higgs
field by a constant amount. In order to construct a Higgs
potential we first must KK decompose the Higgs field
which, together with the quartic terms,
will result in an effective 4-d potential with the desired properties once
$y$ is integrated over. This then allows us to shift the (as we will see) tachyonic
`would-be' zero mode by a constant amount identifying the resulting physical field
with the Higgs.  However, even when brane terms are absent, a
scalar field with a bulk mass in a warped geometry does not
possess a flat mode \cite{Goldberger:1999uk}. In addition, it has been
observed that the mass eigenvalue of the first mode moves exponentially quickly
from zero to order one as the bulk mass is turned on
\cite{Chang:1999nh,Goldberger:1999uk}, so we expect a vev with non-trivial bulk
profile.

Our procedure will be as follows: we first consider the case of the
free, non-interacting Higgs action and solve the corresponding bulk equations of motion
with the brane mass terms, $\mu_{P,H}^2$, supplying the appropriate boundary
conditions. Using the free Higgs action allows us to perform the KK decomposition.  
In certain regions of the parameter space this leads to a single
light tachyonic scalar mode which we can identify as the unshifted Higgs
field. Next we examine the full potential for this tachyonic mode and perform
the usual shift in the field making direct connection with the SM.

The truncated Higgs action (\ie, ignoring gauge and self-couplings
as well as other
particles such as the Goldstone bosons so that we 
take $\Phi \to \phi$) $S_{trunc}$ is given by 
\begin{equation}
S_{trunc}=\int d^5x ~\sqrt {-g}\Bigg[(\partial^A\phi)^\dagger (\partial_A
\phi)-m^2\phi^\dagger \phi+{1\over {k}}\phi^\dagger \phi~[\mu_P^2\delta(y)-
\mu_H^2\delta(y-\pi r_c)]\Bigg]\,.
\end{equation}
To scale out dimensional factors and to make contact with our later development
we define $m^2=20k^2\xi$ and $\mu_{P,H}^2=16k^2\xi\beta_{P,H}$ since $k$ is the
canonical scale for RS masses. Note that the parameters $\xi, \beta_{P,H}$
are dimensionless and
are expected to be $O(1)$ but may, in principle, be of arbitrary sign. The choice
of these unusual looking factors will be made clear below.

What are we looking for in the $(\xi, \beta_{P,H})$ parameter space? To obtain
EWSB in a manner consistent with the SM our basic criterion is to find
regions where there exists one, and only one, TeV scale
tachyonic mode that we can identify with the SM Higgs. The remaining Higgs KK
tower fields must also be normal, \ie, non-tachyonic.
Clearly, if {\it none} of the three mass terms in the action
above are tachyonic no light tachyon will occur in the free Higgs spectrum.

To go further, we must obtain the relevant expressions for the Higgs KK masses
and wavefunctions so we let $\phi \to \sum_n \phi_n(x) \chi_n(y)$.
Substituting this expression into the action above and following the usual
KK decomposition procedure leads to the
equation of motion for the Higgs KK wavefunctions, $\chi_n$:
\begin{equation}
\partial_y\Big(e^{-4\sigma}\partial_y \chi_n\Big)-m^2e^{-4\sigma}\chi_n+
{1\over{k}}e^{-4\sigma}[\mu_P^2\delta(y)-\mu_H^2\delta(y-\pi r_c)]\chi_n
+m_n^2e^{-2\sigma}\chi_n=0\,,
\end{equation}
with $m_n$ being the Higgs KK mass eigenvalues. The solutions are of the
familiar form{\cite{Goldberger:1999uk}}
\begin{equation}
\chi_n={e^{2\sigma}\over {N_n}} \zeta_\nu \Big(x_n\epsilon e^\sigma\Big)\,,
\end{equation}
with $N_n \propto k^{-1/2}$ a normalization factor, $m_n=x_nk\epsilon$, $\epsilon=
e^{-\pi kr_c}\simeq 10^{-16}$, $\nu^2=4+m^2/k^2=4+20\xi$, and
$\zeta_\nu=J_\nu+\kappa_n Y_\nu$,
being the usual Bessel functions. While the $\kappa_n$ are determined by
the boundary conditions on the Planck brane and are generally very small,
of order $\epsilon^{2\nu}$, the values of the $x_n$ are determined from
the boundary condition on the TeV brane. Explicitly one finds
\begin{equation}
-\kappa_n={{\Big[2\Big(1+{\mu_P^2\over{4k^2}}\Big)-\nu \Big]J_\nu(x_n\epsilon)
+x_n\epsilon J_{\nu-1}(x_n\epsilon)}\over {\Big[2\Big(1+{\mu_P^2\over{4k^2}}\Big)-
\nu \Big]Y_\nu(x_n\epsilon)+x_n\epsilon Y_{\nu-1}(x_n\epsilon)}}\,,
\end{equation}
while the $x_n$ roots can be obtained from
\begin{equation}
\Bigg[2\Big(1+{\mu_H^2\over{4k^2}}\Big)-\nu \Bigg]\zeta_\nu(x_n)
+x_n\zeta_{\nu-1}(x_n)=0\,.
\end{equation}
Since the $\kappa_n$ are quite small the values of the parameters $\beta_P$
and, correspondingly, $\mu_P^2$, are generally numerically
irrelevant to the analysis that we will perform below (even for the case
of $\nu=0$ provided $\beta_P$ is $O(1)$). Given the remaining two parameters,
there are four possible sign choices to consider
and we need to explore the solutions
of the equations above in all these cases.
However, we find that if the combination $\xi \beta_H >0$
($\mu_H^2>0$) then either no tachyon exists or that
the resulting Higgs vev is Planck scale, independent
of the sign of $m^2$. This corresponds to an $x_n$ root of the equation above
whose value is of order
$\epsilon^{-1}$; recall that we are seeking a single tachyonic root of order
unity since $m_n=x_nk\epsilon$ and we expect that $k\epsilon$ to be at most
of order a few hundred GeV in order to solve the hierarchy. Clearly,
we must instead choose the parameters such that $\mu_H^2, \xi\beta_H <0$.
In this case the two remaining regions,
\begin{eqnarray}
(I)& & \xi>0, ~\beta_H<0\nonumber \\
(II)& & \xi<0, ~\beta_H>0\nonumber \,
\end{eqnarray}
are found to be quite distinct since in the former case $\nu^2 >0$ is
guaranteed and thus $\nu$ is always real.

What conditions are necessary in these two regions in order to find only
one, single $O(1)$ tachyonic root, $x_T$?
Our first goal is to  find {\it at least} one such root which is $O(1)$.
Let us assume that $\nu$ is real (and positive
without loss of generality). Analytically, the best way to find a
tachyonic root which is $O(1)$ is to look for the conditions on the parameters
necessary to obtain a zero-mode and then to perturb around these. The root
equation, Eq.(8), above already tells us that if
the term in the square brackets is zero
then $x_n\zeta_{\nu-1}(x_n)=0$, implying a root $x_n=0$; thus if
$\nu=2+8\xi\beta_H=0$ we obtain a zero mode. Similarly, if we set the term in the
square bracket equal to $-2\nu$ and use the Bessel function identity,
$2\nu \zeta_\nu(z)=z[\zeta_{\nu+1}(z)-\zeta_{\nu-1}(z)]$, we obtain
$x_n\zeta_{\nu+1}(x_n)=0$, which again has a zero root, and implies
$-\nu=2+8\xi\beta_H=0$. Since $\nu \geq 0$, by hypothesis, these two equations
lead directly to a pair of bounds on $\xi$ which are necessary to satisfy in
order to obtain $O(1)$ tachyonic
roots: we first obtain the bound $\xi \geq(\leq)\xi_1$ in region I(II).
Recalling, in addition, that $\nu^2=4+20\xi=(2+8\xi\beta_H)^2\geq 0$, yields
a second constraint $\xi \geq(\leq)\xi_2$ in region I(II). The $\xi_{1,2}$ are
given by the expressions
\begin{eqnarray}
\xi_1&=&-{1\over {4\beta_H}}\nonumber \\
\xi_2&=&{{5-8\beta_H}\over {16\beta_H^2}}\,.
\end{eqnarray}
Observe that in region I, $\nu$ is always real since $\xi \geq 0$ there.
Furthermore, if these constraints are satisfied we
find there exists only a single $O(1)$ tachyonic root, provided that
$\nu$ is real. Interestingly, if these two constraints
are not satisfied one finds no tachyons whatsoever so that EWSB cannot occur
in those parameter space regions.

What do these constraints look like in regions I and II?
Fig.1 shows all of the constraints in the $\xi-\beta_H$ plane for both regions I
and II. Note that in region II, since $\xi <0$, to maintain our assumption that
$\nu$ is real requires that $\xi \geq -0.2$ as is also shown in the figure. In
region I, since it is always true that $\xi_2>\xi_1$, only the requirement that
$\xi>\xi_2$ as a function of $\beta_H$ is relevant and the allowed region is
rather simple. This is not the case in region II where $\xi_1=\xi_2$ when
$\beta_H=5/4$. In fact with the assumption that $\nu$ is real we are forced
to have $\beta_H \geq 5/4$ in region II since there are now three simultaneous
constraints to satisfy. What if we give up this $\nu^2 >0$ assumption
which populates
the dominant part of region II? When $\xi <-0.2$ then $\nu$ is purely
imaginary and one finds a multitude of $O(1)$ and larger
tachyonic roots, thus violating our
requirements that only one such root should exist. In addition, because $\nu$
is imaginary the factor $\epsilon^{2\nu}$ in $\kappa_n$ becomes a rapidly varying
phase leading to extreme parameter sensitivity. This is apparently a
pathological
regime where it is not obvious that any consistent EWSB
scenario can be constructed. Thus from now on we will ignore the possibility that
$\nu$ may be imaginary.

Fig.2 shows the possible values of the
tachyonic root, $x_T$, as a function of $\xi$ for a wide range of
$\beta_H$ values in both allowed regions. We see that for a significant range of
the $\xi$ and $\beta_H$ parameters that indeed $x_T$ is of order unity.
It is important to observe that the value $\beta_H=1$ is outside of the region II
allowed range.
Further restrictions on the parameter space will occur when we consider the gauge
boson masses in more detail in the next section.
\begin{figure}[htbp]
\centerline{
\includegraphics[width=9.5cm,angle=-90]{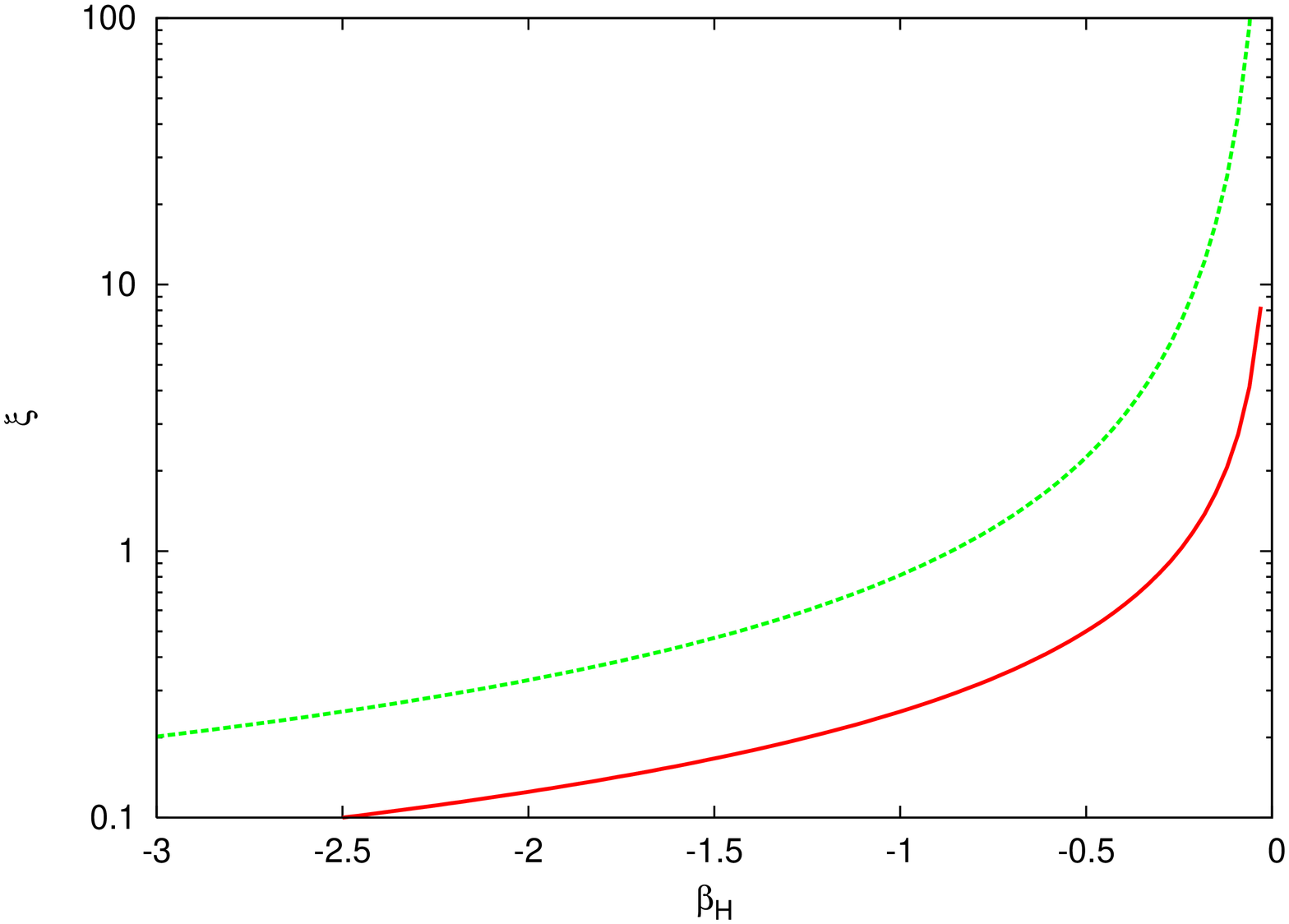}}
\centerline{
\includegraphics[width=9.5cm,angle=-90]{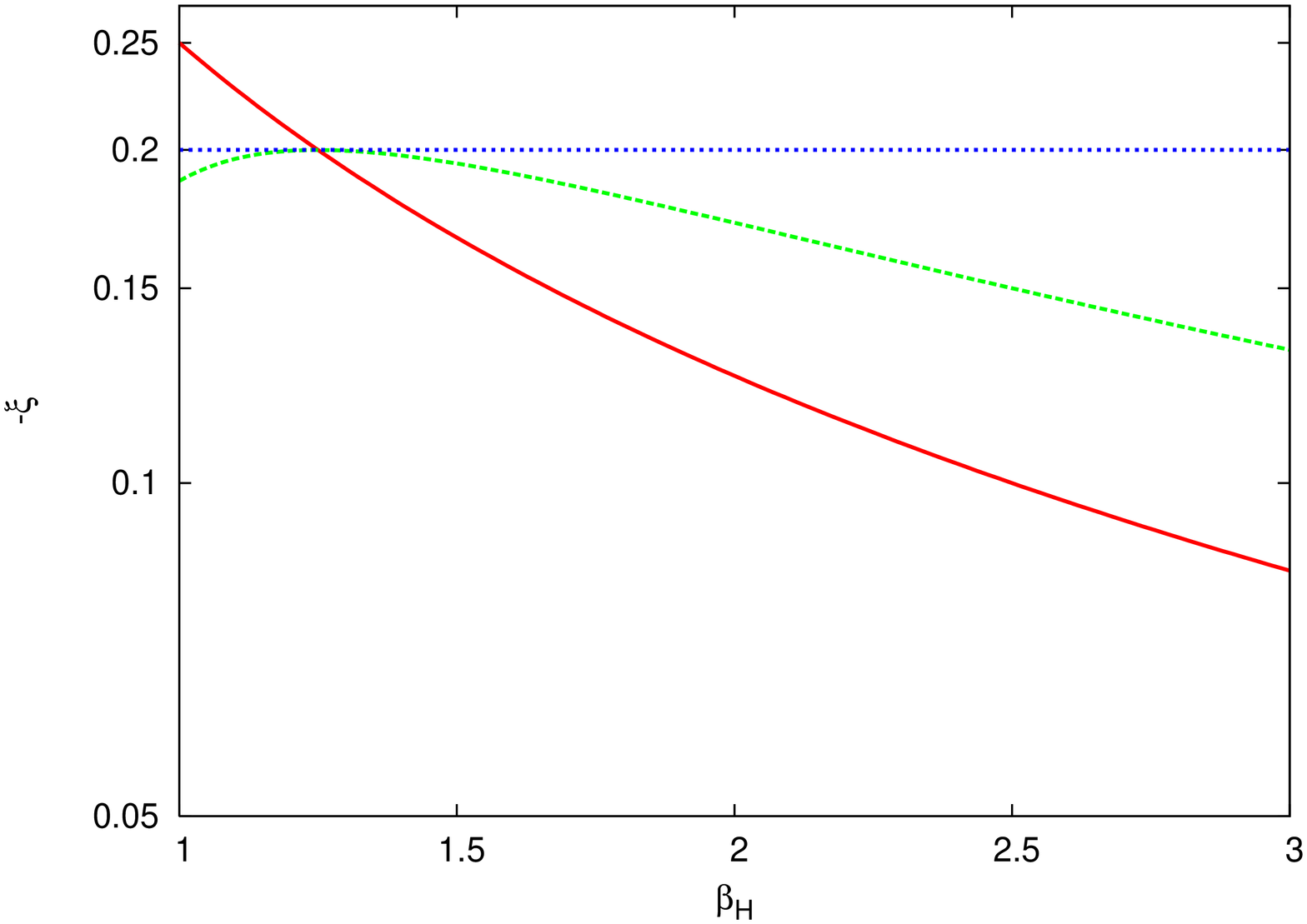}}
\vspace*{0.1cm}
\caption{Allowed regions in the $\xi-\beta_H$ plane for bulk Higgs induced
EWSB in region I (top) and region II (bottom). The lower bound
$\xi\geq -0.2$ (dotted blue) that insures $\nu^2 \geq 0$ in region II is also shown
in addition to both constraints $\xi_1$ (dashed green) and $\xi_2$ (in solid red). The
allowed region lies between the blue and green curves in region II and above
the green curve in region I.}
\label{fig1}
\end{figure}
\begin{figure}[htbp]
\centerline{
\includegraphics[width=9.5cm,angle=-90]{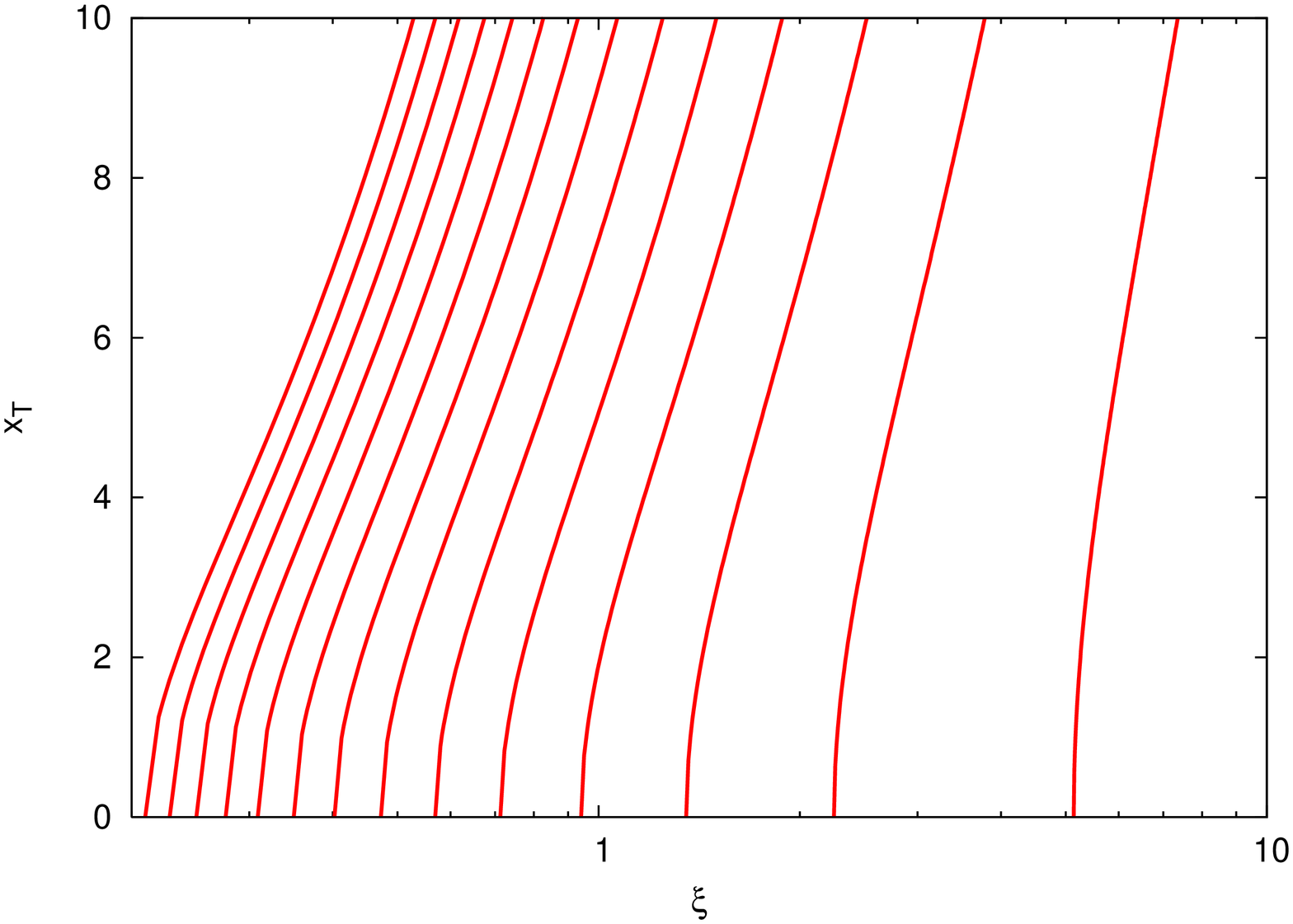}}
\centerline{
\includegraphics[width=9.5cm,angle=-90]{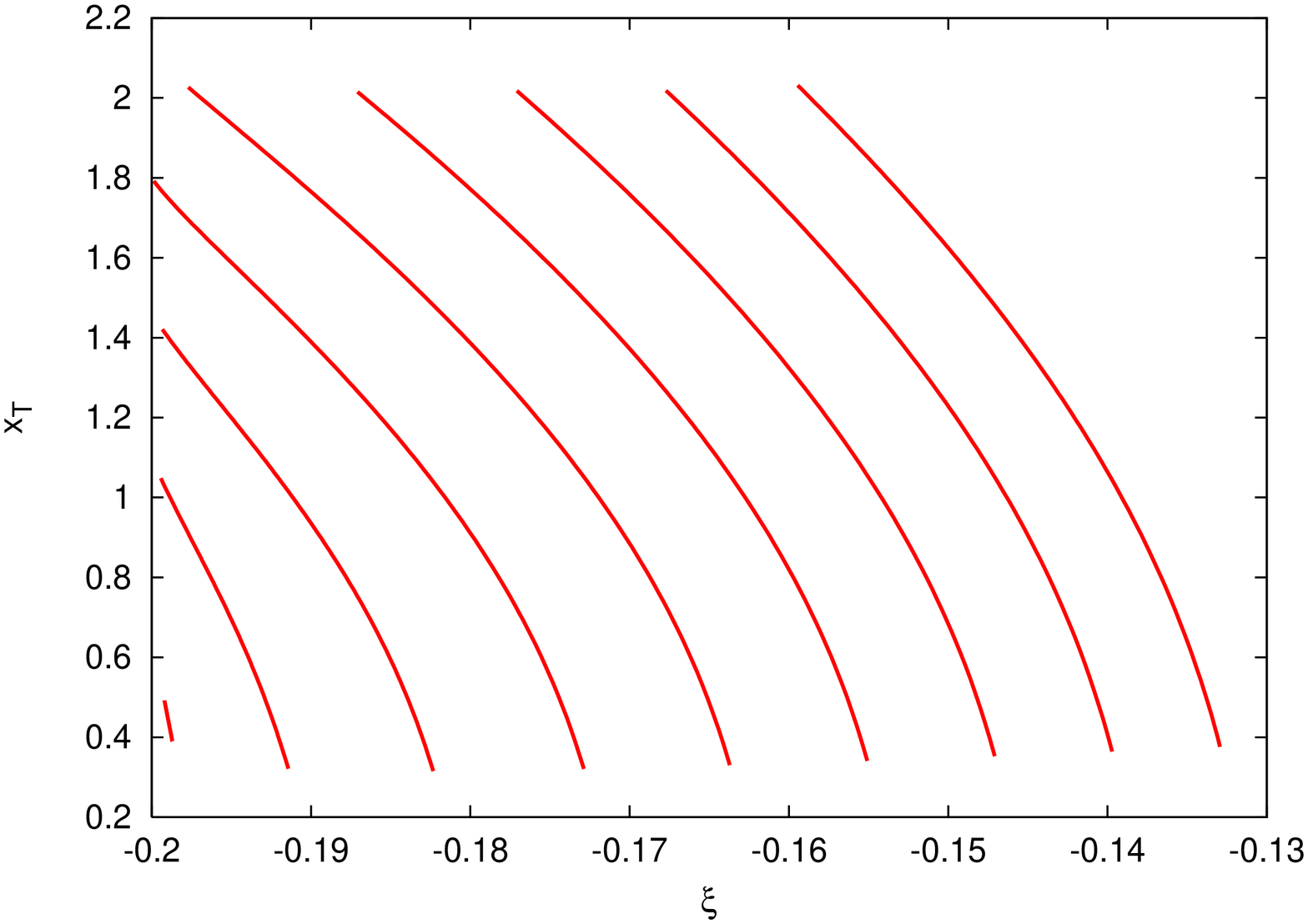}}
\vspace*{0.1cm}
\caption{Tachyon roots as a function of $\xi$ for different $\beta_H$. In
region I (top), with $\beta_H$ ranging from $-0.3$ to $-2.9$, going from right to
left in steps of 0.2 and with $\beta_H$ values from 1.4 to 3.0, going left
to right, in steps of 0.2 for region II (bottom).}
\label{fig2}
\end{figure}

Noting the mass of the Higgs tachyonic field and inserting
the Higgs KK decomposition $\phi =\sum_n \phi_n(x) \chi_n(y)=\phi_T\chi_T+...$,
back into the action, integrating over $y$ and
extracting only the resulting potential for the 4-d part of the
tachyonic field, $\phi_T$, yields the familiar expression
\begin{equation}
V=-m_T^2\phi_T^\dagger \phi_T+\lambda_4 (\phi_T^\dagger \phi_T)^2\,,
\end{equation}
where
\begin{equation}
\lambda_4=\frac{\lambda_H}{k^2}\int ~dy~\sqrt {-g}~\chi_T^4(y)\delta(y-\pi r_c)\,,
\label{lam4}
\end{equation}
so that we can shift the field $\phi_T\to (v+H)/\sqrt {2}$ as usual.
Here, we note that the contribution from the Planck brane quartic
term is exponentially suppressed and is ignored.
There are
several things to note about this: ($i$) the physical `SM' Higgs mass is given
by $m_H=\sqrt {2} m_T=\sqrt {2}x_T k\epsilon$ and is fixed relative to the
masses of all the other KK states and ($ii$) although the tachyonic $\phi$ mass
is on the TeV brane, and possibly in the bulk as well,
the Higgs vev now has a non-trivial profile in the
5-d bulk given by $\chi_T/\sqrt{2}$. As we can easily see this function is very
highly peaked near the TeV brane since $\chi_T \sim e^{(2+\nu)\sigma}$.
This is why we `see' a TeV scale vev and Higgs boson mass and also why we avoid
some of the previous problems with placing the Higgs field in the RS
bulk{\cite{Chang:1999nh}}.  Note that the overall shape of the Higgs profile is
{\it adjustable} by varying the parameters $\xi$ and $\beta_H$.
($iii$) It is important to observe that the vev, $v$, in the expression
above is of the same scale as
that of the SM Higgs, $\sim 246$ GeV, and is {\it not} Planckian.

It is important to note that the presence of the quartic terms in the 4-d effective
potential that leads to a positive mass squared for the Higgs produces a small shift
in the Higgs wavefunction relative to the vev profile. In lowest order of perturbation
theory we can write this shift as
\begin{equation}
\chi_H(y)=\chi_T(y)-{3\over {2}}\sum_{n=1} {x_T^2\over {(x_T^2+x_n^2)}}R_{1n}\chi_n(y)
\,,
\end{equation}
where $R_{1n}$ is just the ratio of wavefunctions
\begin{equation}
R_{1n}= \chi_n(\pi r_c)/\chi_T (\pi r_c)
\end{equation}
which is less than unity as we will be demonstrated below. Here $x_n$ are the roots
corresponding to the Higgs KK excitations. Since as we will later see, $(x_n/x_T)^2
\grtsim 100$, the above correction is at the level of a per cent and thus will be
neglected in the discussion that follows, \ie, $\chi_H=\chi_T$, will be assumed from
now on.

As the addition of the quartic term in the potential leads to a shift in the mass of
the physical Higgs making it non-tachyonic, other terms in the 4-d quartic potential,
$\sim \phi_T^2 \phi_n^2$, lead to small modifications in the masses of the Higgs KK states
when we shift the field $\phi_T\to (v+H)/\sqrt {2}$. A short calculation leads to the
result
\begin{equation}
\Delta m_n^2=
{3\over {4}}m_H^2 R_{1n}^2\,.
\end{equation}
We will return to this point further below where we we will see that this shift is
at most on the per cent level since $R_{1n}<1$ and the KK masses are large compared
to $m_H$.

\section{Gauge Boson Masses}

The masses of the SM gauge bosons are, as usual, generated via the kinetic terms
in $S_{Higgs}$. Unlike the usual RS-type scenario, the Higgs vev is no longer
restricted to the TeV brane but has a profile, $\chi_T$, in the extra dimension.
To extract the gauge boson mass terms from $S_{Higgs}$ we can perform the
standard KK decomposition and combine these terms with those obtained from the
relevant pieces of $S_{gauge}$ obtaining the gauge boson wavefunctions and
KK mass spectra. To be specific, let us consider the case of the $W$ boson.
Suppressing Lorentz indices, we employ the KK decomposition $W=\sum_n W_n(x)
f_n(y)$ and obtain the following equation for the gauge KK states:
\begin{equation}
\partial_y\Big(e^{-2\sigma}\partial_y f_n\Big)-{1\over {4}} g_5^2v^2\chi_T^2
e^{-2\sigma}f_n+m_n^2f_n=0\,,
\end{equation}
where $g_5$ is the 5-d $SU(2)_L$ gauge coupling. It is important to notice that
($i$) the 4-d Higgs TeV-scale vev, $v$ appears here, {\it not} a 5-d vev
and ($ii$) the tachyonic Higgs profile is present in the mass generating term.
Here, we have included the back-reaction of of the bulk Higgs profile on
the gauge field KK equation of motion.  Neglecting this back-reaction
would have yielded the ``free" field wavefunctions for the gauge KK fields,
starting with a massless mode.  To proceed, one would then integrate over
the bulk degrees of freedom, using the free wavefunctions, and obtain a
4-d mass matrix for the $W$ KK modes.  To perform a diagonalization, one
would have to truncate this mass matrix, in practice.  The off-diagonal
elements of this matrix are not small compared to the mass squared of the
lightest $W$ mode, and hence we do not expect this procedure to yield
accurate results, using a modest truncation.  Since we are going to
compare the properties of the lightest $W$ mode with those required from
precision electroweak data, we choose to consider the exact equation above
instead.

Now we would like to solve this equation to obtain the $W$ KK spectrum and
wavefunctions; however, due to the presence of $\chi_T^2(y)$, which is a
combination of Bessel functions of imaginary argument,
an analytic solution is not obtainable. We may,
however, obtain a fairly good approximate solution by remembering that $\chi_T$
is strongly peaked at the TeV brane in which case $\chi_T^2 \to \epsilon^{-2}
\delta(y-\pi r_c)$. One can thus solve the resulting equation exactly, but then we
would have no idea how good our approximation is. We will refer to this as the
`$\delta$-function limit'. The
overall validity of this approximation will certainly improve for larger $\xi$
since then $\chi_T^2 \sim e^{2(\nu+2)\sigma}$ becomes even more sharply peaked
near the TeV brane in this case.

To obtain a better approximation,
we performed the following calculation: we let $\chi_T ^2 \to
\lambda \epsilon^{-2}\delta(y-\pi r_c)$, where $\lambda$ here is a free
parameter, and solved this equation analytically for the usual KK spectrum and
eigenfunctions. We then treat the difference
\begin{equation}
V_{pt}={1\over {4}} g_5^2v^2\Big[\chi_T^2e^{-2\sigma}
-\lambda\delta(y-\pi r_c)\Big]\,,
\end{equation}
as a first order `perturbing potential' and calculate the elements of the
$W$ mass squared matrix for the
KK states. Lastly, we vary the parameter $\lambda$ until this
mass-squared matrix is as diagonal as possible. In particular, we are specifically
interested in making the off-diagonal elements in the first row and column be
as small as possible as these influence the $W$ mass via mixing. Our expectation
is that $\lambda$
will be near unity if our approximation is valid. Performing
this analysis for several choices of the input parameters tells us that indeed
$\lambda \simeq 1.1-1.2$ in region I where $\nu$ is large and, in region II,
$\lambda \simeq 1.1-1.3$ where $\nu$ is smaller. Thus the $\delta$-function
limit is
a reasonable approximation in both regions and can be improved upon by choosing
a $\lambda$ in the above range. We now have gotten the $W$ KK masses and
eigenfunctions to a very good approximation. Obtaining the KK masses \etc ~for
the other SM gauge fields can be done in a parallel manner. To connect with the
4-d SM we must also relate the 5-d coupling, $g_5$, to the usual SM $g$. If
the SM fermions are localized near either brane this can be
done by defining the weak coupling as that between the brane fermions and the
would-be $W$ zero mode, \ie, $g^2=g_5^2f_0(y=0,\pi r_c)^2$.
If BKT's for the gauge fields are present, this definition is easily modified
to include such effects{\cite{Davoudiasl:2002ua}}.

\begin{figure}[htbp]
\centerline{
\includegraphics[width=8.5cm,angle=90]{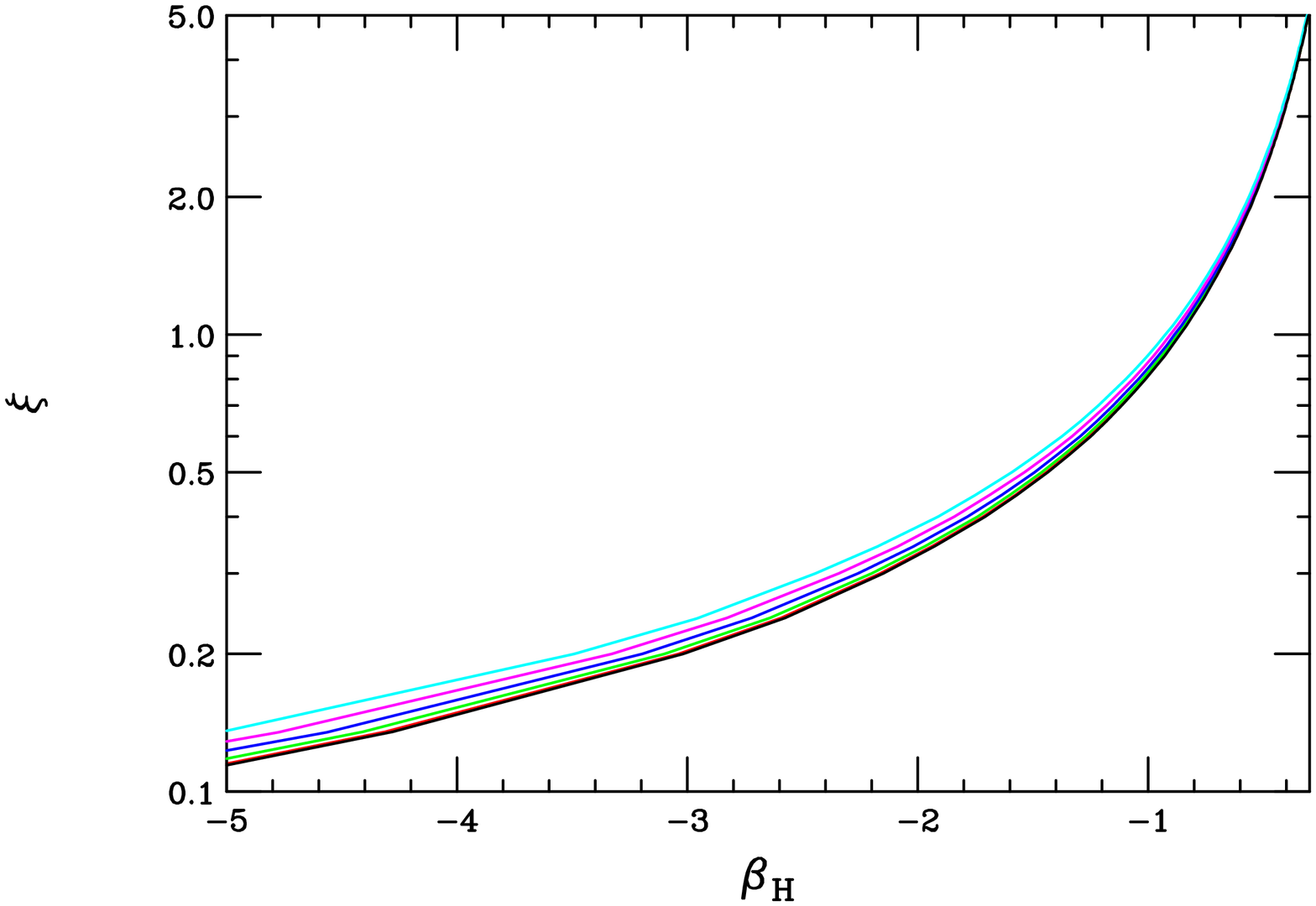}}
\vspace*{0.5cm}
\centerline{
\includegraphics[width=8.5cm,angle=90]{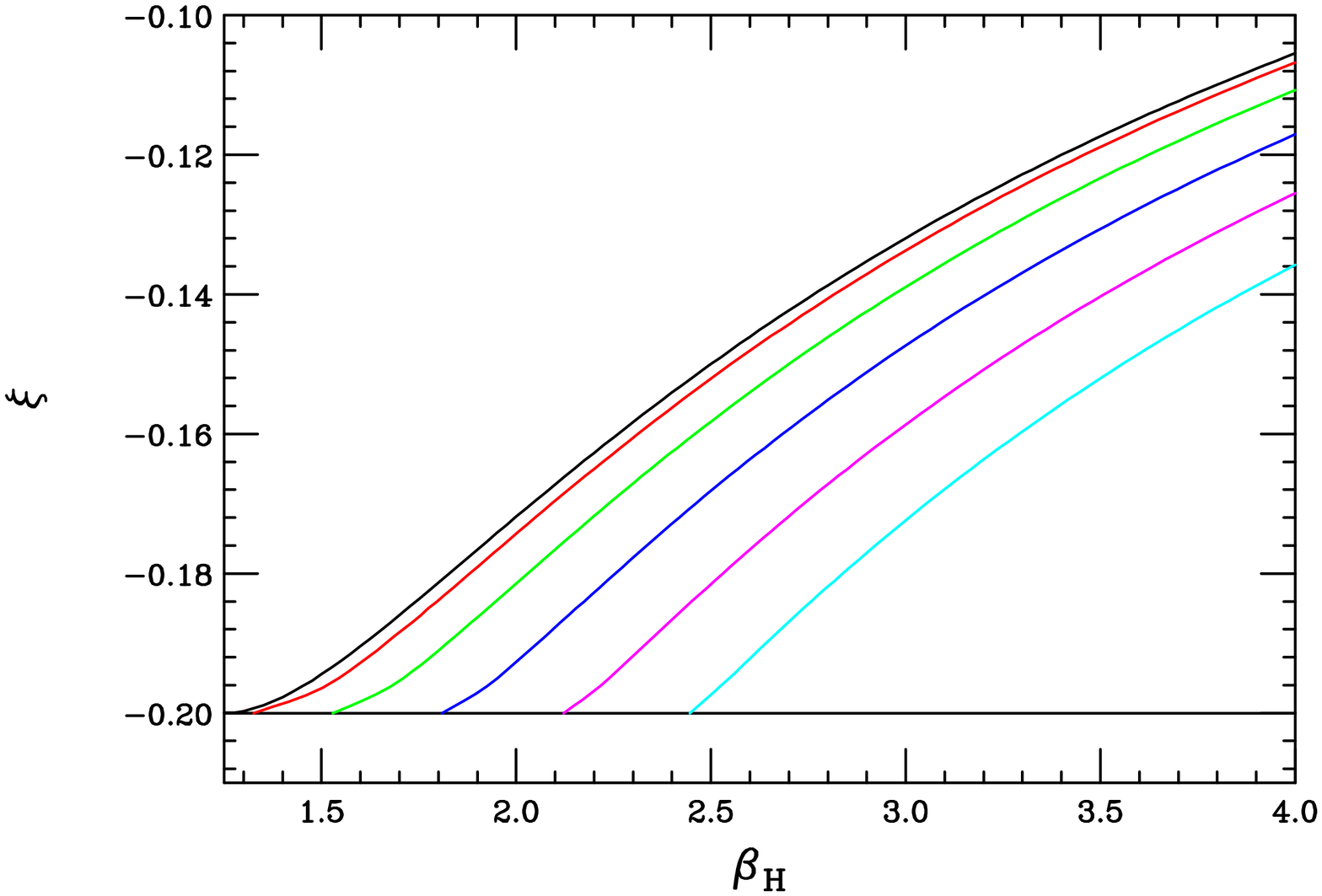}}
\vspace*{0.1cm}
\caption{The curves correspond to fixed tachyon root
values, $x_T=0$ (black) to 2.5 (cyan), in steps of 0.5,
 as functions of $\xi$ and $\beta_H$ in regions I (top) and II (bottom).}
\label{fig3}
\end{figure}

From this analysis we can extract the `roots' corresponding to the mass of,
\eg, the $W$ boson in the usual manner, \ie, $M_W=x_W k\epsilon$. We typically find
that $x_W \simeq 0.20-0.25$ from which we may infer that $k\epsilon \sim 350$ GeV.
Using this result we would conclude that $x_T=1$ corresponds to a Higgs mass
of $\sim 500$ GeV so that this suggests that somewhat smaller values of
$x_T \sim 0.5$ might be considered more favorable. From Fig.~\ref{fig3} we
see that it is easy to obtain such values
in both regions I and II as long as we live near the $\xi_{1,2}$ boundaries.
How are the $W$ and $Z$ masses correlated in this model?  Without extending the
gauge group in the bulk to $SU(2)_L\times SU(2)_R\times U(1)_{B-L}$, as would be
done in a more realistic model{\cite{Csaki:2003dt,Agashe:2003zs}}, there is
no custodial symmetry present
to insure the validity of the SM $M_W=M_Z \cos \theta_w$ relationship,
\ie, $\rho=1$. As discussed in Ref.{\cite {new}},
if $v/(k\epsilon)\lsim 1$ then the gauge KK mass scales approximately linearly in
$v$ and we can obtain $\rho \simeq 1$. This relation will become more exact
as $v/(k\epsilon)$ becomes smaller. However, when $v/(k\epsilon)$ becomes
very large then the generated gauge boson
mass becomes independent of $v$ and we would thus obtain $M_W=M_Z$, a gross violation
of custodial symmetry. Since we sit between the two extremes, $v\sim k\epsilon$,
we certainly would expect such violations to be of some significance
in the present case. Sampling the model parameter space we do indeed find
sizable deviations of $\rho$ from unity due to custodial symmetry violation. Here
we make use of the approximate gauge boson mass matrix calculations described above,
and obtain deviations which are of order $5-10\%$, some of which may arise from
the approximations we have made in obtaining the roots and wavefunctions.
These deviations can be cured through an extension of the SM gauge group which
enforces custodial symmetry as discussed above.

\section{Gravity-Induced Electroweak Symmetry Breaking}

As an example of EWSB with a bulk Higgs we now turn to the special case where
EWSB is triggered by gravity.

\subsection{Gravity Action}

In this case we need to consider a generalized
version of the usual RS gravitational action which augments the above action:
\begin{equation}
S=S_{bulk}+S_{branes}+S_{Higgs}+S_{gauge}\,,
\end{equation}
where now the various terms are given by
\begin{eqnarray}
S_{bulk}&=&\int d^5x ~\sqrt {-g} \Bigg[{M^3\over {2}} R-\Lambda_b+
{\alpha M\over {2}} \Big(R^2-4R_{AB}R^{AB}+R_{ABCD}R^{ABCD}\Big)
\Bigg]\nonumber \\
S_{branes}&=&\sum_{branes} \int d^5x ~\sqrt {-g} \Bigg[{M^3\over {k}}
\gamma_i R_4-\Lambda_i - \frac{\lambda_i}{k^2}(\Phi^\dagger \Phi)^2
\Bigg]\delta(y-y_i)\nonumber \\
S_{Higgs}&=&\int d^5x ~\sqrt {-g}\Bigg[(D^A\Phi)^\dagger (D_A \Phi)
+\xi R\Phi^\dagger \Phi\Bigg]\,,
\end{eqnarray}
and $S_{gauge}$ is as above.
In these expressions $M$ is the 5-d Planck or fundamental scale, $R$ is the
usual 5-d Ricci curvature with $\Lambda_b$ being the bulk cosmological
constant, all of which are
familiar from the usual RS model. Generally, $k$ and $M$ are
expected not to be very different in magnitude in order to avoid introducing
another hierarchy problem. The piece of $S_{bulk}$
which is quadratic in the curvature is the GBT with $\alpha$ being a dimensionless
parameter which we take to be of indefinite sign
and roughly of $O(1)$. The $\gamma_i$ are
graviton BKT's for the TeV ($i=\pi$) and Planck ($i=0$) branes and $R_4$ are
the 4-d Ricci scalars derived from the corresponding induced metrics on both
branes. The $\Lambda_i$ are the usual RS brane tensions. The term
$\xi R\Phi^\dagger \Phi$ in $S_{Higgs}$ is the {\it bulk} Higgs-curvature
mixing term with $\xi$ soon to be identified with the parameter introduced in
Section II.
This is the most general form allowed for the gravitational action in the RS
model framework.

There are no bare Higgs mass terms of any kind in this action as they are
assumed to be generated from the Ricci scalar $R$.
In order to generate the Higgs mass terms let us briefly
recall that in the usual RS
scenario, the 5-d Ricci scalar obtains a `vev', \ie, the RS metric itself
produces a background value for $R$. Employing the conventional RS relationships
between the bulk and brane tensions, Einstein's equations lead to
$\langle R\rangle=-20k^2+16k[\delta(y)-\delta(y-\pi r_c)]$ so that with, \eg,
$\xi>0$, $\langle R\rangle$ induces a positive mass squared for the Higgs
in the bulk and on the TeV brane and a negative mass squared on the Planck
brane. (Note that the original RS model
requires that $\beta_H=\beta_P=1$.) This would lead to
a 4-d theory where the Higgs vev is of Planck scale as we found previously
so that we must instead choose $\xi<0$. However, in that case we saw that
$\beta_H \geq 5/4$ is required to obtain EWSB. Thus, unless we alter the RS
model in some way we cannot achieve gravity-induced EWSB.

The important role played by the GBT in this action is to maintain the general
properties of the RS model while allowing  an extension of the
parameter space, \ie, to $\beta_H \neq 1$, for this scenario to be
phenomenologically successful. As was recognized long ago by Kim,
Kyae and Lee and subsequently discussed by other authors{\cite{Kim:2000pz}}, the
GBT allows us to modify the relationship between the tensions of the TeV and
Planck branes and the other RS parameters which adds additional flexibility
in the model.  In particular one now finds that
\begin{equation}
\Lambda_{Planck}=-\Lambda_{TeV}=6kM^3\Big(1- {4\alpha k^2 \over {3M^2}}
\Big)\equiv 6kM^3\beta_H\,,
\end{equation}
so that the gravity induced effective free Higgs action is just
\begin{equation}
S_{eff}=\int d^5x ~\sqrt {-g}\Bigg[(\partial^A\Phi)^\dagger (\partial_A
\Phi)-m^2(\Phi^\dagger \Phi)+{\mu^2\over {k}}\Phi^\dagger \Phi[\delta(y)-
\delta(y-\pi r_c)]\Bigg]\,,
\end{equation}
and we identify $m^2=20k^2\xi$ and $\mu^2=16k^2\xi\beta_H$ as above in Section
II. Given our
previous general analysis, the allowed parameter space for gravity-induced EWSB
is {\it already} known.

\subsection{Classical Stability}

With the addition of gravity to our original action,
the $\xi$ term leads to a number of new effects,
in particular mixing among the whole Higgs tower and radion fields. To address
these issues requires several steps: First, we must extract from the action
the kinetic term for the radion and see that it is properly normalized.
Although this is straightforward it is non-trivial for the case at hand. Here we
follow the work of Csaki, Graesser and Kribs(CGK){\cite{Giudice:2000av}} and
expand the metric as in their Eq.(3.2):
\begin{equation}
ds^2=e^{-2\sigma-2F}\eta_{\mu\nu}dx^\mu dx^\nu -[1+2F]^2 dy^2\,.
\end{equation}
We write their quantity $F$ as $e^{2\sigma}r_0(x)$.
Next we insert this metric into the $S_{bulk}$, $S_{branes}$ and
$S_{Higgs}$ terms in the action, perform a series expansion keeping
terms only through second order in the derivatives of $r_0$, and integrate over
$y$ dropping terms which are subleading in $\epsilon$. This results in a kinetic
term for $r_0$ of the form
\begin{equation}
{6M^3\over {k}}(\partial r_0)^2N_r^2\,,
\end{equation}
where $N_r^2$ sums over several distinct contributions:
\begin{equation}
N_r^2=\Big(1-4\alpha{k^2\over {M^2}}\Big)(1-2\Omega_\pi)\,,
\end{equation}
with
\begin{equation}
\Omega_{0,\pi}\equiv {{4\alpha k^2/M^2\pm \gamma_{0,\pi}}\over{1-4\alpha k^2
/M^2}}\,.
\end{equation}
Note  that $N_r^2$ contains the usual contributions from the Ricci scalar as well
as those from both the GBT and BKT's. This
quantity must be positive definite to avoid ghosts in the radion
sector and this can be much more easily accomplished
in region II where $\beta_H$ is positive and $\alpha$ is negative.

As has been discussed by several authors{\cite{Kim:2000pz}}, the
presence of the BKT's and GBT in the RS model can result
in the graviton and/or radion field becoming a ghost and the possible presence
of a physical tachyon in the graviton spectrum; avoiding these problems
constrains the model parameters as we saw in the case of $N_r^2$ above.
Furthermore, eliminating the possibility of tachyons in the gravity sector
requires{\cite{Kim:2000pz}} that the parameter $\Omega_\pi<0$. Given this condition
$N_r^2>0$ is
automatically satisfied in region II while seemingly very difficult to satisfy in
region I.

To make contact with TeV scale physics we must relate $M^3/k$ to
$\mpl^2$; recall that in the simple RS scenario $M^3/k=\mpl^2$ neglecting
terms of order $\epsilon^2$.
In turns out that in the case of the graviton, the requirement that the norm of
this field also be positive definite, \ie, be ghost-free, is the same
as requiring that $\mpl^2$ be positive definite when expressed in terms of the
parameters in the action. Writing

\begin{equation}
\mpl^2={M^3\over {k}}N_g^2\,,
\end{equation}
a straightforward calculation{\cite {me}} leads to the relation
\begin{equation}
N_g^2=\Big(1-4\alpha{k^2\over {M^2}}\Big)(1+2\Omega_0)+
{{\xi kv^2}\over {M^3\epsilon^2}}>0\,,
\end{equation}
where again $O(\epsilon^2)$ terms have been neglected and we have taken the
$\delta$-function approximation for simplicity in the last term since it is
$\sim 0.01$ or less. It is difficult for us to obtain $N_{r,g}^2>0$ in region
I simultaneously. In region II, to obtain
$N_g^2>0$ we only require that $\Omega_0>-1/2$
which is a rather mild constraint. Note that if both $N_{r,g}^2>0$ then the
graviton KK's also have positive definite norms. A curious observation
is that there exists a small but finite region of the parameter space where
no graviton brane terms are required to obtain both $N_{r,g}^2>0$; in such a
case $\Omega_0=\Omega_\pi \grtsim -1/2$.
Finally, putting all these pieces together we can write the normalized radion
field, $r$, as
\begin{equation}
r_0=r{\epsilon^2\over {\sqrt {6}\Lambda_\pi}} {N_g\over {N_r}},
\end{equation}
where $\Lambda_\pi=\mpl \epsilon$.
Observe that in the original RS model both $N_{r,g}=1$; the ratio
$N=N_g/N_r$ will occur frequently in the expressions below.

Thus we conclude that we
can obtain a working model of gravity-induced EWSB provided we live in region II.
As shown in Figs. 1 and 3, typical values of $\xi$ in region II are
close to the conformal value $\xi=-3/16$.

\subsection{Higgs-Radion Mixing}

Since the normalized form of the radion field is now known, we can return to the
action and extract out pieces which are quadratic in the scalars, perform the
associated KK decompositions (ignoring for now the KK expansions of the
Goldstone bosons which do not mix with the radion as we will see
below) and integrate over $y$ leaving us with the 4-d effective Lagrangian
(a sum over $n$ is implied)
\begin{eqnarray}
{\cal L}&=&-{1\over {2}}H_n \Box H_n-{1\over {2}}m_{H_n}^2H_n^2
-{1\over {2}}m_r^2 r^2\nonumber \\
&+& \xi \gamma A_nH_n\Box r-{1\over {2}}\Big[1+B\xi\gamma^2\Big]r\Box r\,,
\end{eqnarray}
which is analogous to that obtained by CGK{\cite{Giudice:2000av}} in their
Eq.(10.12); here $\gamma={v\over {\sqrt {6}\Lambda_\pi}}\simeq 0.01-0.05$.
Defining $n={T,i}$ with `$T$' labeling the mode that gets a vev, the
coefficients are given by
\begin{eqnarray}
A_T&=&2\epsilon^2N\int ~dy~\chi_T^2 \simeq 2N \nonumber \\
A_i&=&2\epsilon^2N\int ~dy~\chi_T \chi_i \simeq 0 \nonumber \\
B&=&-6\epsilon^4N^2\int ~dy~e^{2k|y|}\chi_T^2 \simeq -6N^2 \,.
\end{eqnarray}
where the approximate results hold in the $\delta$-function limit. Employing a
scan of the parameter space we find that these approximations hold to better than
$\simeq 50\%$; the true values tend to be somewhat below those given by the
$\delta$-function approximation, \eg, $A_T\simeq 1.3(1.7)$ and
$-B\simeq 3.4-4.1(4.8-5.5)$ in region II(I).
Here we make some note of the factor `2' appearing in the definition of $A_n$
above;
this value is a result of the mixing terms being bulk operators with the full
5-d Ricci scalar. In the standard case of wall Higgs fields, $R\to R_{ind}$ and
`2' $\to$ `6' and the results of CGK are recovered.
The kinetic mixing can be removed as usual by suitable field redefinitions:
\begin{eqnarray}
H_n&\to& H_n'+\xi \gamma\beta A_nr'\nonumber\\
r&\to&\beta r'\,,
\end{eqnarray}
with
\begin{equation}
\beta^{-2}=1+B\xi\gamma^2-\xi^2\gamma^2A_n^2\,,
\end{equation}
where a sum on the index $n=(T,i)$ is now understood.
Demanding that $\beta^{-2}$ be positive places another constraint on our model
parameters, in particular, we obtain a bound on the parameter $\xi$:
\begin{eqnarray}
{B\over {2A_n^2}}\Bigg[1+\Bigg(1+{{4A_n^2}\over {\gamma^2B^2}}\Bigg)^{1/2}\Bigg]
\leq \xi \leq {B\over {2A_n^2}}\Bigg[1-\Bigg(1+{{4A_n^2}\over {\gamma^2B^2}}
\Bigg)^{1/2}\Bigg]\,.
\end{eqnarray}
Given our parameter space this bound is easily satisfied.
Mass mixing remains but this can be easily read off from ${\cal L}$ above and
follows the usual course{\cite{Giudice:2000av}}.

\section{Experimental Signatures}

In this section we will discuss some of the phenomenological features of our
model as well as how our framework may be experimentally tested.

In a fully realistic model, all SM fields should live in the bulk. The 5-d
fermion masses will be chosen so that the overlap of the would-be zero mode
with $\chi_T$ produces the correct Yukawa couplings. Some mechanism also needs
to be introduced to protect the $\rho$ parameter, such as enlarging the gauge
group to
$SU(2)_L \times SU(2)_R \times U(1)_{B-L}${\cite{Csaki:2003dt,Agashe:2003zs}}.
It is known that, in this
type of model with a Higgs on the IR brane, there is a tension between
producing the correct top mass and not causing too large a shift in the $Zb\bar
b$ coupling. With the non-trivial profile $\chi_T$ this problem should be
softened, since the left-handed top (and hence the left handed bottom) field
can be moved closer to the Planck brane.

\subsection{Higgs KK Excitations}

Our framework describes a wide class of potential models with a new feature:
the fundamental SM Higgs field is in the RS bulk and thus has a scalar KK
excitation spectrum{\cite{Kehagias:2005be}}.
This is a completely new feature previously not considered within the RS model
structure.
Once the values of $\beta_H$ and $\xi$ are specified in our model so are  the
ratios of the masses of the Higgs KK's to that of the usual Higgs. (We will ignore
radion mixing in this discussion for simplicity since it is likely to have
little influence on the Higgs KK excitations.) Exploring this parameter space
one generically finds that the
first Higgs KK excitation has a mass $\sim 30-100(15-30)$ times
larger than the $W$ in region I(II), with the largest ratios
obtained for $\xi$ near its boundary value.
This suggests a mass range for the lightest Higgs KK state of $\grtsim 1-1.5$
TeV and is generally found to be more massive than the first graviton KK excitation.
To address this point, Fig. 4 shows the root values for
the first Higgs KK as a function of $\beta_H$ for different values of $\xi$ in
both regions. Here we see that almost all of the time the Higgs KK
is more massive than the first graviton
KK. It is important to remember that the masses of all the Higgs KK excitations
are fixed once the values of $\xi$ and $\beta_H$ are known.

We remind the reader that the physical masses of the Higgs KK excitations are slightly
shifted from the values given by the root equation, $x_nk\epsilon$, due to the
presence of the quartic term in the potential. Given the expression for $\Delta m_n^2$
above we obtain
\begin{equation}
m_n^2=[x_n^2+{3\over {2}}x_T^2R_{1n}^2](k\epsilon)^2\,.
\end{equation}
Since $R_{1n}$ are found to be $<1$ and for almost all parameter space
regions of interest $x_T/x_n \leq 0.1$ it is clear that the magnitudes of these
shifts in the Higgs KK spectrum are at most at the per cent level and can be neglected
in our discussion below.

\begin{figure}[htbp]
\centerline{
\includegraphics[width=8.5cm,angle=90]{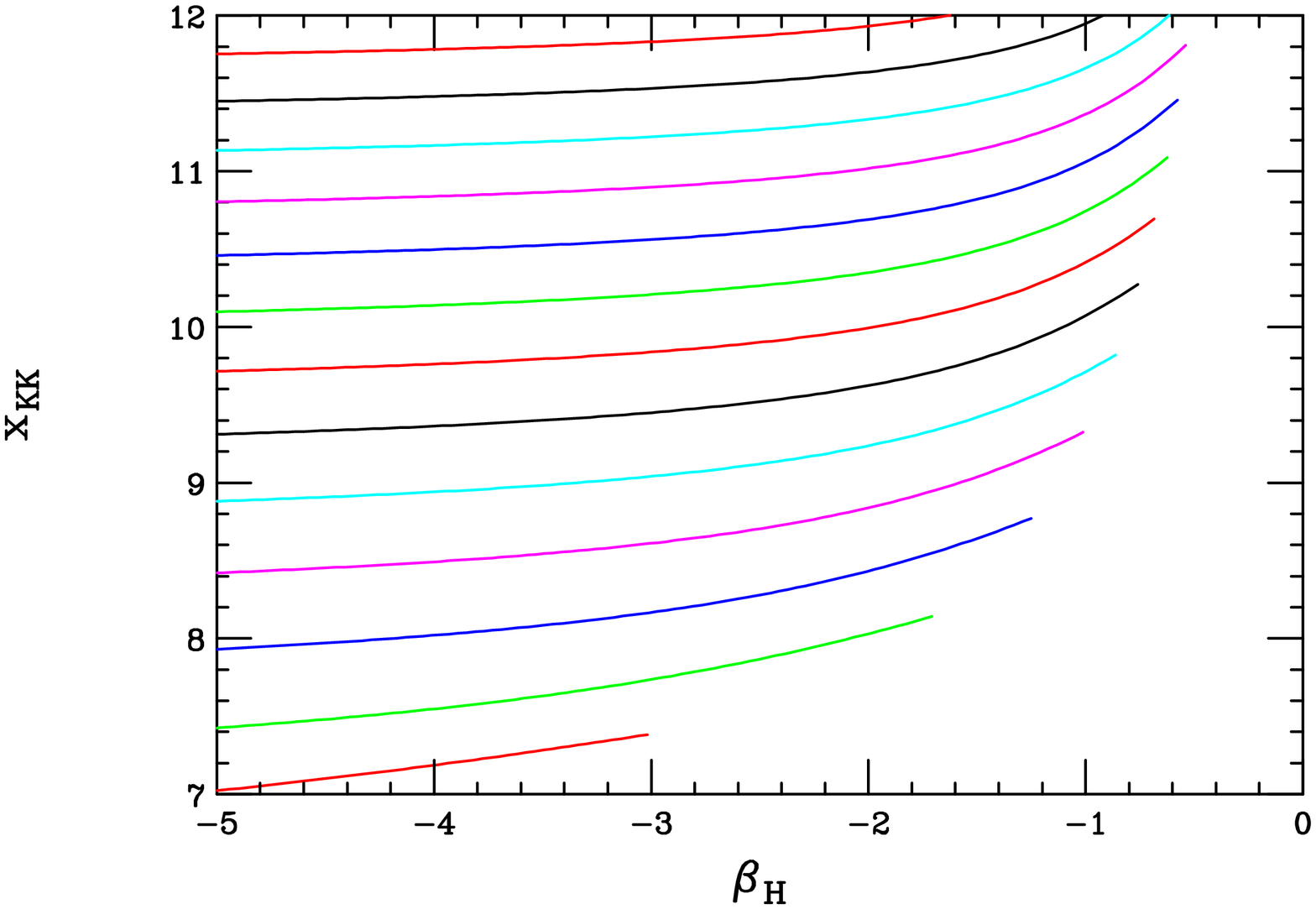}}
\vspace*{0.3cm}
\centerline{
\includegraphics[width=8.5cm,angle=90]{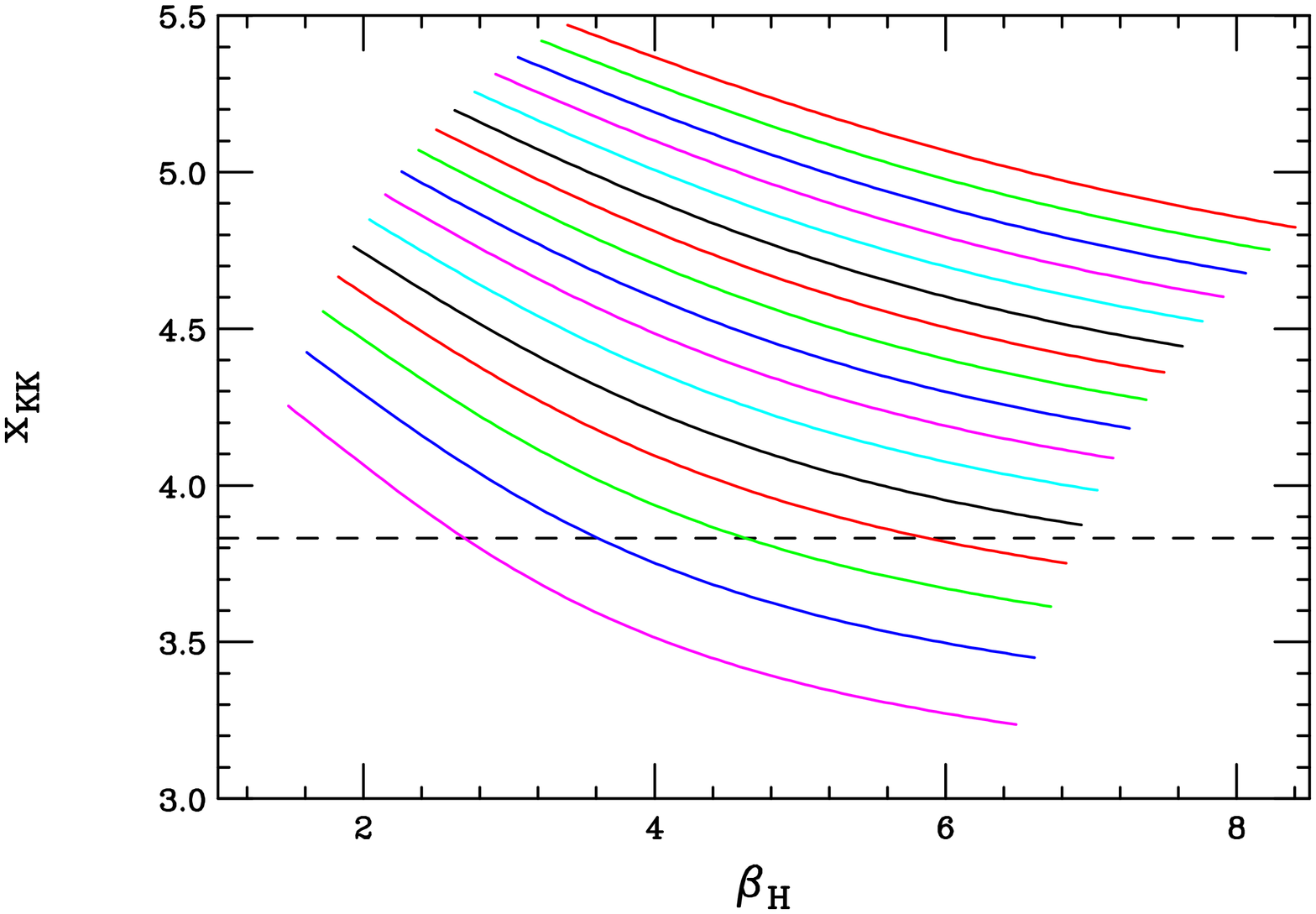}}
\vspace*{0.1cm}
\caption{Roots for the first Higgs KK excitation as a function of $\beta_H$ for
different values of $\xi$ in regions I (top) and II (bottom).
In region I, $\xi=0.2$
is the lowest curve with $\xi$ increasing by 0.2 for each subsequent curve. The
curves are cut off on the right hand side by the $\xi$ constraints.
In region II, from top to bottom $\xi$ runs from $-0.120$ to $-0.195$ in
steps of 0.005. The largest possible
value for the first graviton KK root consistent with the constraint
$\Omega_\pi \leq 0$ found in the case of gravity induced breaking is shown as
the dashed line.}
\label{fig4}
\end{figure}

Can such a KK state be observed at planned colliders? Unless the couplings are very
large the Higgs KK states seem to be beyond the range of the LHC but may be
produced by $Z$ bremsstrahlung at multi-TeV linear colliders such as CLIC.
To address this issue we must ask how a Higgs KK couples to SM fields.
To obtain the $H_nWW$-type couplings we return to the Higgs kinetic term in
$S_{Higgs}$ and extract out the term which is bilinear in the $W$ field yet linear
in the Higgs. Then we perform the usual KK decomposition for each field and
extract the relevant coupling which we can express as an integral over $y$:
\begin{equation}
S_{eff}= \int d^4x~\sum_n H_nW_\mu^\dagger W^\mu ~{1\over {2}}g_5^2v \int~dy~
e^{-2\sigma}\chi_T \chi_n f_0^2\,.
\end{equation}
Interestingly we see that if the $W$ wave-function were flat the integral would
vanish by orthogonality suggesting a small coupling. However, we also know that
that the most significant deviation in flatness for this
gauge boson wavefunction occurs near the TeV brane where $\chi_T$ is largest.
To set the scale for this coupling and to make the connection with the conventional
SM $WWH$ coupling strength, $g^{SM}_{WWH}=g^2v/2$, we note
that if we take the light fermions to be localized near the Planck brane, we can
define $g_5^2=2\pi r_cg^2$. We can then perform the integral above and scale the
resulting $WWH_1$ effective coupling to the SM value.
A scan of the parameter space in both regions I and II indicates that the coupling
of $H_1$ to $WW$ is substantially suppressed in comparison to the SM $WWH$ value.
We find that $g_{WWH_1}/g^{SM}_{WWH} \simeq 0.02-0.13$ with the larger values
appearing in region II. With these rather small couplings it is clear that a large
integrated luminosity will be necessary to detect the $H_1$ KK state if it is
produced by gauge boson fusion or radiated off a gauge boson leg.

It is interesting to note that a similar expression to the above with $\chi_n \to
\chi_T$ allows us to calculate the usual Higgs coupling to the $W$. Ordinarily one
would imagine that this is purely proportional to $M_W^2$ as in the SM. Here, due to
the back-reaction of the gauge boson wavefunction this is no longer the case.
This can be see from Eq.(15) by setting $n=0$, multiplying by $f_0$, integrating
over $y$ and solving for $m_0^2=M_W^2$. Using orthonormality, integrating by-parts
and employing the usual
$\partial_y f_0=0$ boundary conditions on both branes gives
\begin{equation}
M_W^2={1\over {4}}g_5^2v^2\int ~dy ~e^{-2\sigma}f_0^2\chi_T^2 +\int ~dy~e^{-2\sigma}
(\partial_y f_0)^2\,,
\end{equation}
where the first term on the right is directly due to the Higgs boson vev
while the second corresponds
to back-reaction. Here we can see explicitly that these two components are of the
same sign; from this we can conclude that the $HWW$ coupling in this model is less
than in the SM. This shift in the coupling of the Higgs to gauge bosons can be
significant and can be measured at the LHC/ILC. As in the case of $H_1$ above,
we can scale the $WWH$ effective coupling by the corresponding SM value.
A numerical scan of the parameter
space in regions I and II indicates that $g_{WWH}/g^{SM}_{WWH}\simeq 0.45-0.70$,
with the larger values obtained in region II. This large shift in coupling
strength will be easily observable at the ILC and likely also to be seen at the
LHC. Of course the appearance of a
suppression of the $WWH$ coupling in the RS model framework is
not limited to the present scenario and is thus not a unique feature of the present
scheme{\cite {Agashe:2003zs}}.

The Higgs self-interactions as well as the Higgs KK's coupling to SM Higgs
can be obtained through the quartic self-couplings
appearing in $S_{Higgs}$. To this end, we remind ourselves that $\Phi^\dagger
\Phi$ can be written as
\begin{equation}
\Phi^\dagger \Phi= {1\over {2}}(v+H)^2\chi_T^2 +(v+H)\chi_T\sum_n\chi_n H_n+
{1\over {2}}\sum_{n,m}\chi_n\chi_m H_nH_m +G^+G^-+{1\over {2}}G^0G^0\,,
\end{equation}
where here
the $G$'s are the non-KK expanded 5-d Goldstone particles. We first observe
that there are no trilinear or quartic couplings between a {\it single} Goldstone
KK and the SM Higgs; however, such terms for the couplings of two Higgs to a
single Higgs KK excitation do exist. To
evaluate these terms we first observe that we can immediately relate
$\lambda_H$ to the SM quartic coupling, $\lambda_{SM}=m_H^2/(2v^2)$, via
the integral
\begin{equation}
\lambda_{SM}=\frac{\lambda_H}{k^2}\int~dy~\sqrt{-g}\chi_T^4 \,\delta(y-\pi r_c)\,.
\end{equation}
Thus the $H_nHH$ and $H_nHHH$ couplings can be obtained from the action
\begin{equation}
S_{eff}'=\sum_n \lambda_{eff_n} \int~d^4x~(3vH^2+H^3)H_n\,,
\end{equation}
where
\begin{equation}
\lambda_{eff_n}=\lambda_{SM}R_{1n}.
\end{equation}
In a similar manner, one can calculate the equivalent of the SM Higgs trilinear
coupling; one finds that one recovers the SM expression when the definition of
$\lambda_{5\Phi}$ in terms of $\lambda_{SM}$ given above is employed.
A sampling of the model parameter space indicates that $\lambda_{eff_1}/\lambda_{SM}
\simeq 0.65-0.91$ so that the $H_1 \to HH$ decay mode branching
fraction will be sizable.

Given the properties of the $H_1$ KK state it is likely that this particle can
be most easily produced in $gg$-fusion or in $\gamma\gamma$ collisions which
can proceed through top quark loops. The reason for this is that with the top in
the RS bulk it is likely that the $H_1t\bar t$ coupling will remain reasonably
strong. Another possibility would thus be the associated production of
an $H_1$ together with $t \bar t$.

\subsection{Direct Tests}

In the case of gravity-induced EWSB there are strong correlations between the
Higgs and gravity sectors.
Although there are many parameters in the model, they are already constrained
by the analysis above and they are all likely to be accessible through
future collider experiment. This implies that the scenario becomes
overconstrained and the model structure can be directly tested. As we will see,
such measurements will require LHC/ILC and likely a multi-TeV $e^+e^-$
collider for detailed studies of both the scalar and graviton sector. The
reason for this is clear: we need precision measurements and the masses of some
of the important states can be in excess of 1 TeV.

Let us begin with the gravity sector: We assume that the first three graviton KK
excitations, $G_{i=1,2,3}$, are accessible and that their properties can be
determined in detail at $e^+e^-$ colliders.
A measurement of the {\it relative} KK masses gives us $\Omega_\pi$ as this
mass ratio
depends only on this single parameter and any individual mass then provides
the quantity $k\epsilon$. A measurement of the ratio of partial widths
for the same final sate, \eg, $\Gamma(G_2\to \mu^+\mu^-)/\Gamma(G_1\to
\mu^+\mu^-)$, yields a determination of a  combination of the parameters
$\Omega_{0,\pi}$ while an {\it overall} width measurement yields us
$\Lambda_\pi$. The graviton KK spectrum is such that $G_3\to 2G_1$ is
kinematically allowed{\cite{Davoudiasl:2001uj}} and the possibility of
studying such decays in detail at
$e^+e^-$  colliders has already been discussed in the
literature{\cite{Battaglia:2001id}}. The rate and angular
distribution for this processes is sensitive to both the existence of BKT's
as well as the GBT{\cite{Zwiebach:1985uq}} allowing
us to separate these two contributions once $\Lambda_\pi$ is known. However, such
precision measurements of $G_3$ are likely to require a multi-TeV linear collider.
When $\Omega_\pi \lsim -0.28$, the decay $G_2\to 2G_1$ is also allowed with
different contributing
weights arising from the BKT's and GBT contributions. In either case, it is
likely that the graviton sector alone will tell us $\alpha$,
$\gamma_{0,\pi}$, $\Lambda_\pi$, $k/M$ and hence $\beta_H$.
As one can see, a combination
of just these measurements is already rather restrictive and may be sufficient to
confirm or exclude the present model.

Now for the scalar sector: We assume that the two lightest states, which are
mixtures of radion with the Higgs, can both be observed so
that their masses and couplings can be well determined. From these data it will
be possible to reconstruct the mixing matrix which provides for us the `weak'
eigenstate mass parameters as well as a determinations of $\xi$. Given $\xi$, the
Higgs and first KK Higgs masses, the value of $\beta_H$ can again be extracted
and compared with that obtained from the gravity sector. A confirmation of the
model is obtained if the two values agree.

\section{Conclusions}

The consistency of having a fundamental 5-d Higgs doublet in the RS
geometry was studied in this work.  We showed that by assigning appropriate
bulk and brane masses for the 5-d ``Off-the-Wall'' Higgs, one can achieve a
realistic 4-d picture of EWSB, without fine-tuning.  In our construct, the SM
Higgs is the lightest KK mode of the bulk doublet which in the 4-d reduction
has a tachyonic mass, leading to EWSB. Since all of the SM fields are now in
the RS bulk, this scenario represents an example of a {\it warped} Universal
Extra Dimension.

Previous attempts at EWSB with a bulk Higgs field in the RS geometry were
plagued by extreme fine-tuning and phenomenological problems{\cite{Chang:1999nh}}.
Nearly all such models had considered endowing the bulk Higgs with a 5-d
constant vev which yielded massive SM gauge bosons in 5 dimensions.  However,
in our framework, EWSB involves only one KK mode of the bulk Higgs and
is hence localized.  The resulting Higgs vev has a bulk-profile that is
nearly identical to that of the physical SM Higgs.
Although the consistency of the bulk Higgs scenario
imposes non-trivial constraints on our framework,
we observe that a realistic 4-d phenomenology can be achieved for a range
of parameters.

A particularly interesting realization of the Off-the-Wall Higgs EWSB employs the
gravitational sector of the RS model to provide the necessary bulk
and brane mass scales.  These scales
are then related to the 5-d curvature of the RS
geometry.  Consequently, new relations and constraints among the parameters of the
Higgs and gravitational sectors are obtained.  In this ``gravity-induced" EWSB
scenario, the higher curvature Gauss-Bonnet terms
play an important role.  The coupling of the
gravity and Higgs sectors as well as the introduction of
the higher derivative terms affect
the classical stability of the RS geometry.  We find that one of the two regions
of parameter space is favored by these stability considerations.
As we have shown, typical values of $\xi$ in region II are
close to the conformal value $\xi=-3/16$.

We outlined the experimental tests and phenomenological features of the Off-the-Wall
Higgs and gravity-induced EWSB in our work.  A novel feature of our general
bulk scenario is the existence of an adjustable
5-d profile of the Higgs vev which can provide a new tool
for model building. Generically, we also expect
the appearance of TeV scale Higgs KK
modes.  In the case of gravity-induced EWSB, observation of radion-Higgs mixing,
together with measurements of the Higgs and graviton KK modes can provide
tests of this mechanism at future colliders; TeV-scale $e^+ e^-$ colliders are
well-suited for this purpose.

In summary, we have presented a consistent framework for placing the Higgs
doublet in the RS geometry bulk and achieving EWSB at low energies.  The possibility
that 5-d gravity drives 4-d EWSB is considered and shown to be a viable option.
With the Higgs residing in the RS bulk, the entire SM can be thought of as a 5-d
theory and it also becomes feasible to think of bulk gravity as the cause of EWSB.
This provides a cohesive 5-d picture of all the known forces of nature that is
both predictive and free of large hierarchies.

{\it Note added}: After this paper was essentially
completed, Ref.{\cite{Flachi:2000wi}}
was brought to our attention where the possibility of using the bulk curvature in
the RS model to generate the Higgs vev was also considered. As discussed in this
work, the parameter space point employed by these authors leads to multiple tachyons
in the scalar spectrum and is plagued by extreme parameter sensitivity.

\acknowledgments

H. Davoudiasl would like to thank  K. Agashe,
D. Chung and F. Petriello for discussions
related to this paper. This work was supported in part by the
U.S.~Department of Energy under grant DE-FG02-95ER40896 and contract
DE-AC02-76SF0051.  The research of H.D. is also
supported in part by the P.A.M. Dirac Fellowship, awarded by the
Department of Physics at the University of Wisconsin-Madison. The authors would
also like to thank the Aspen Center of Physics for their hospitality while
this work was being completed.

\end{document}